\documentclass{sf2a-conf2025}
\usepackage{graphicx}
\usepackage{hyperref}
\usepackage[]{natbib}  
\usepackage{epstopdf}

\def\BibTeX{{\rm B\kern-.05em{\sc i\kern-.025em b}\kern-.08em
    T\kern-.1667em\lower.7ex\hbox{E}\kern-.125emX}}
\bibpunct{(}{)}{;}{a}{}{,}  %%%%%%%%%%%%%  A&A bibliography style
\begin{document}

\TitreGlobal{SF2A 2025}

%%-----------------------------------------------------------------
%%      the top matter
%%

\title{Probing binary supermassive black hole in quasars through spectropolarimetry}

\runningtitle{Polarimetric insights into a potential binary SMBH in Mrk 231}

\author{J. Biedermann}\address{Universit\'e de Strasbourg, CNRS, Observatoire astronomique de Strasbourg, UMR 7550, F-67000 Strasbourg, France}

\author{F. Marin$^1$}

%% IF Author3 has the same affiliation than Author1:
%\author{C.\,E. Author3$^1$}

%% IF Author3 has its own affiliation:
%\author{C.\,E. Author3}\address{Dept. of Chess, University of Games, 35101 Las Vegas, Monaco} 

%% IF Author3 has two affiliations, the one of Author1 and a second one:
%\author{C.\,E. Author3$^{1,}$}\address{Dept. of Chess, University of Games, 35101 Las Vegas, Monaco} 

%% Keep this line, even if the page will be settled afterwards.
\setcounter{page}{237}

%%-----------------------------------------------------------------

\maketitle

%%-----------------------------------------------------------------
%%        The abstract
%% 
%%  Warning!  within the abstract:
%%  - do not use macros. 
%%  - do not use commands like: \cite, \citet, \citep ... etc.

\begin{abstract}
Nearly all galaxies are believed to host a supermassive black hole (SMBH) at their center. In the context of galaxy mergers, the SMBHs from the progenitor galaxies are expected to form a bound binary system. Given the high frequency of such mergers throughout cosmic time, binary SMBHs should be relatively common in active galactic nuclei (AGNs). Yet, their direct detection remains a major observational challenge, with most candidates inferred only through indirect signatures. The quasar Markarian 231 (Mrk 231) presents unique ultraviolet (UV) - optical properties, including an atypical drop in the total flux around $\sim$ 2500 \AA \ and a strong wavelength-dependent polarization. In this work, we explore the potential of spectropolarimetry as a tool to detect and characterize SMBH binaries in AGNs. We construct a simple analytical model in order to reproduce the spectral energy distribution (SED) of Mrk 231 taking into account a binary SMBHs scenario and we compute the resulting total and polarized flux. Our goal is to test if a the presence of two distinct sources can reproduce the unique UV-optical properties observed in Mrk 231. Such investigations support the case for future high-sensitivity UV spectropolarimeters in order to probe the inner structure of quasars. In particular, the next-generation instrument POLLUX, proposed for the Habitable Worlds Observatory (HWO), would be particularly suited to provide new insights into AGNs physics and SMBH binaries in quasars.
\end{abstract}

%% Insert the keywords (to appear in the ADS indexing)
%% Keywords must be separated by a comma

%%-----------------------------------------------------------------

\section{Introduction}
%%---------------------

AGNs are among the most luminous and energetic objects in the Universe. In the unified model of AGNs proposed by \citet{Antonucci_1993}, AGN activity is powered by a SMBH, with a mass ranging from millions to billions of solar masses, surrounded by an accretion disk. This disk is responsible for most of the optical and UV emissions observed in AGNs. Surrounding the central engine lies a dusty torus, which can obscure the inner regions depending on the line of sight of the observer. Between the accretion disk and the torus is the broad line region (BLR), composed of dense and rapidly rotating clouds of gas which are responsible of the broad emission lines seen in some AGN spectra. Further out, along the polar axis, resides the narrow-line region (NLR), a structure of less dense gaz, that can extend to kilo-parsec scales, responsible for narrow emission lines. In some AGNs, jets of ionized matter are launched at relativistic speeds from the central engine. While the unified model provides a general picture of the structure of AGNs, there's more and more evidence that, in some cases, the central engine might not have just one, but two SMBHs. These configurations can really change the geometry the central regions, potentially showing up in the spectral and polarization properties of AGNs.

The Markarian 231 (Mrk 231) quasar is a particularly intriguing object in this context. It is one of the brightest ultraluminous infrared galaxies in the nearby Universe that displays unique spectral and polarization properties in the UV-optical range \citep{Thompson_1980}. The total flux shows an abrupt decline from optical to UV, with a pronounced break around 2500 \AA, followed by a slow increase at shorter wavelengths. This behavior contrasts with typical quasars, where the UV flux generally decreases more gradually. In addition, Mrk 231 displays unusually high UV polarization in the same waveband region where we observe the cut-off in the total flux, with the degree of polarization being strongly wavelength dependent, rising from a few percent in the far-UV to around 15\% in the optical, one of the highest levels of polarization observed in non-blazing quasars, accompanied by a corresponding rotation of the polarization angle \citep{Smith_1995, Goodrich_1994, Smith_2004}.

To explain the significant break in total flux, at $\sim$ 2500 \AA, \citet{Yan_2015} proposed that Mrk 231 could host a binary SMBHs system, composed of a more massive primary SMBH, dominating the optical emission, and a less massive secondary SMBH, located in the circumumbinary disk, emitting mainly in the far-UV. This hypothesis is attractive because Mrk 231 is in an advanced phase of galactic fusion, and is therefore a strong candidate to host a binary system of SMBHs \citep{Armus_1994}. In this work, we aim to reproduce the spectral energy distribution (SED) of such a binary system and investigate whether this model can explain the spectroscopic and polarization properties observed in Mrk 231.

\section{Analysis}
%%-------------------------
\subsection{Modeling the Spectral Energy Distribution of of a binary SMBHs system}

We developed a simple analytical model in order to reproduce the SED of a binary SMBHs system. As a first step, we modeled the emission of a single SMBH using the standard accretion disk model described by \citet{shakura_Sunyaev}, where the disk is geometrically thin and optically thick, allowing it to emit locally as a blackbody. To define the outer radius of the disk, we used the Toomre stability criterion, since it indicates that the disk becomes gravitationally unstable when the accretion rate of the system surpass a certain limit, depending mainly on the mass of the SMBH. With these elements, we successfully reproduced the SED of a single SMBH. 

We then extended the model to a binary SMBHs system following the scenario proposed by \citet{Yan_2015}, in which the secondary smaller SMBH is located inside the inner region of the circumbinary disk of the primary bigger SMBH. In this configuration, the secondary SMBH opens a gap in the primary disk and accretes matter from the circumbinary environment, forming a new mini-disk that emits mainly in the far-UV range. To define the geometry of this system, we computed the Hill radius to estimate the inner radius of the primary disk, we used the Roche lobe radius to determine the outer boundary of secondary mini-disk and the Innermost Stable Circular Orbit (ISCO) for the inner radius of the mini-disk. We included those calculations in our model and the sum of the two emission reproduces the SED for a binary SMBH. 

As we wanted to compare our model to Mrk 231 data, we compiled multi-wavelength flux measurement of the quasar with corresponding polarization measurement. These data points are shown in Fig.~\ref{Biedermann_fig1} where the colors indicate the aperture of the instrument. In order to construct the total SED of Mrk 231 across a broad wavelength range, we included the contribution from the main components of the quasar : the host galaxy for the optical emission and the dusty torus in the infrared waveband. To do so, we used templates coming from the Swire library proposed by \citet{Polletta_2007} and the SED simulations of \citet{Siebenmorgen_2001} for the host galaxy and the torus respectively. We sum the four contributions, the primary and secondary SMBH, the host galaxy and the dusty torus, in order to obtain the total SED extending from the far-UV to IR. The combined model is displayed by the red curve in the top-left panel in Fig.~\ref{Biedermann_fig1}.

The model well reproduces the observed shape of Mrk 231 data and well reproduces the breaks visible around $\sim$ 2500 \AA \ too. In our interpretation, this break corresponds to the separation between the inner primary disk and the outer secondary mini-disk of the two SMBHs. The far-UV peaks below $\sim$2500 \AA \ is attributed to the contribution of the secondary smaller SMBH and the contribution beyond this wavelength is attributed to the emission of the primary bigger SMBH.

\subsection{Polarimetric modeling}

Through the orientation of the electric field of light, polarization provides additional constraints on the geometry of AGNs \citep{Marin_2019}. We can describe it using the Stokes formalism, which defines a four dimensional vector, $S=(I, Q, U, V)$, where $I$ is the total intensity, $Q$ and $U$ describe linear polarization, and $V$ the circular polarization. In this study, we focus on linear polarization only since there are no circular polarization measurements for this object. From the Stokes parameters $Q$ and $U$, we derive two key quantities: the degree of polarization $P=\frac{\sqrt{Q^{2}+U^{2}}}{I}$, and the polarization angle $\theta = \frac{1}{2}arctan(\frac{U}{Q})$. The polarized flux is then calculated as the product of the degree of polarization and the total flux. For each component of our SED model, the primary and secondary SMBH disks, the host galaxy, and the dusty torus, we assigned best-fit values of the normalized Stokes parameters : $q_{i}=Q_{i}/I_{i}$ and $u_{i}=U_{i}/I_{i}$. These values were chosen based on observational constraints and physical expectations. We then summed the Stokes flux of all components to compute the polarization signal of the system.

\begin{figure}[ht!]
 \centering
 \includegraphics[width=1\textwidth,clip]{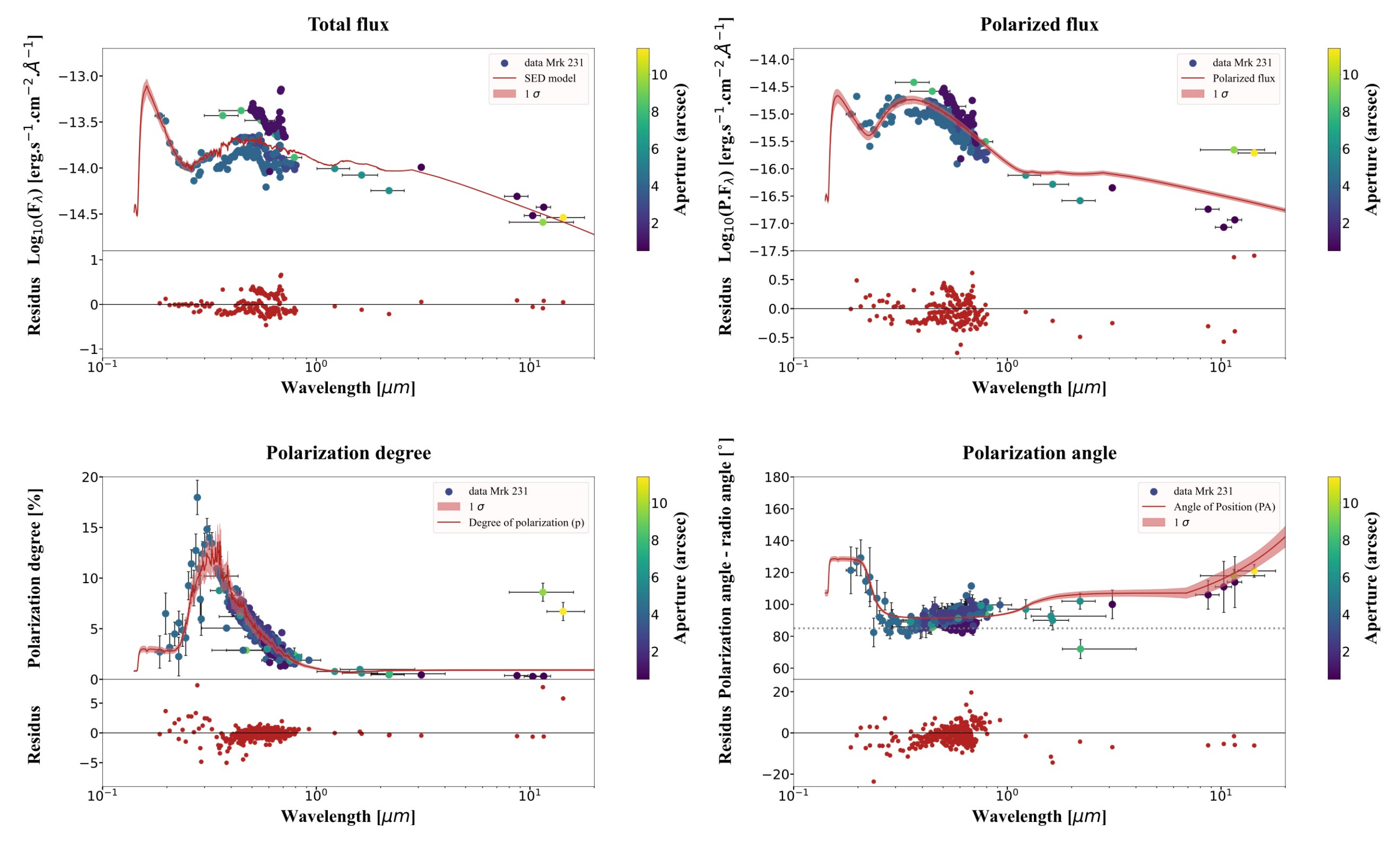}    
%% Note the ABSENCE of the extension .pdf  !
  \caption{\textit{Top left panel}: Mrk 231 total flux as a function of wavelength. \textit{Top right panel}: Polarized flux, established from the multiplication of the flux and the polarization degree. \textit{Bottom left panel}: Degree of polarization. \textit{Bottom right panel}: Polarization position angle minus the radio position angle \citep{Biedermann_2025}.}
  \label{Biedermann_fig1}
\end{figure}

The resulting degree of polarization, polarization angle, and polarized flux are shown in Fig.~\ref{Biedermann_fig1} (bottom-left, bottom-right, and top-right panels, respectively) and our best-fit polarization parameter for each component of the model are summarized in Table~\ref{param_pol_final}. Our model successfully reproduces the observed variation of the degree of polarization in Mrk 231. In particular, the primary SMBH dominates the polarized emission in the optical, allowing the model to reach the observed polarization level of $\sim$15\%. We assumed zero polarization value for the host galaxy component, as stellar emission is typically unpolarized. In the UV domain, the model predicts a smooth variation in the polarization angle, particularly near the spectral break at $\sim$ 2500 \AA. This behavior is interpreted as the result of the transition between the emission of the secondary SMBH in the far-UV and the primary SMBH in the optical, highlighting the contribution of the binary SMBHs system onto the observed polarization features. In addition, the polarized flux predicted by our model displays a second bump in the far-UV, corresponding to the emission from the secondary, less massive SMBH. Altogether, the model is able to simultaneously reproduce three independent observables (the total flux, the degree of polarization, and the polarization angle), which support the binary SMBHs interpretation in order to explain the spectropolarimetric properties of Mrk 231.

\begin{table}
    \centering
    \caption{Table of parameters and best-fit values for the polarization model.}
    \begin{tabular}{ c c c }
        \\
        \hline
        Component & q & u\\ 
        \hline
        Primary SMBH & -0.2254 $\pm$ 0.0282 & -0.0475 $\pm$ 0.0009 \\
        Secondary SMBH & -0.0012 $\pm$ 0.0016 & -0.0310 $\pm$ 0.0036 \\
        Torus & -0.0066 $\pm$ 0.0009 & -0.0064 $\pm$ 0.0003 \\
        Host galaxy & 0 & 0 \\
        \hline
    \end{tabular}
    %\tablefoot{}
    \label{param_pol_final}
\end{table}

%%
%% Example of single figure
%% PLEASE ONLY PROVIDE PDF FIGURES
%%

However, the current far-UV data remain poorly sampled, which prevents us from confirming the presence of the predicted bump with high confidence. Future high-sensitivity spectropolarimetric observations in this wavelength range will be crucial to test the robustness of the presence of this second bump and to further assess the validity of the binary SMBHs hypothesis in the core of Mrk 231.

\section{Conclusions}
%%--------------------
Mrk 231 displays unique UV spectropolarimetric signatures. We constructed a basic analytical model based on a simple binary SMBHs configuration. Our results support the binary SMBHs scenario for Mrk 231 by successfully reproducing three independent observables: the total flux, polarization degree, and polarization angle. The model predicts a second bump in the far‐UV polarized flux corresponding to the contribution of the smaller secondary SMBH, which may constitutes a polarization signature for binary SMBHs system in quasars. However, the lack of far‐UV spectropolarimetric data prevents a direct confirmation. This work opens promising perspective for future space‐based UV spectropolarimetry to probe the inner structure of quasars and test the binary SMBHs scenario in AGNs using UV spectropolarimetry. Especially, the next generation instrument POLLUX, a high resolution UV spectropolarimeter proposed by an European consortium for the Habitable Worlds Observatory (HWO) could provide key constraints \citep{Neiner_2025}. Covering the 100 $nm$ et 1.5 $\mu m$ range with unprecedented spectral resolution (R$\sim$120000 in the far-UV from 100 to 120 $nm$), POLLUX will allow detailed measurements of polarized light in the far-UV range, where binary SMBHs polarization signatures are most prominent. This will open the way to identify polarization signatures of binary SMBHs and characterize their properties.

% Optional acknowledgements
% -------------------------
\begin{acknowledgements}
Thank you to the organizers of the SF2A 2025!
\end{acknowledgements}

\bibliographystyle{aa}  % A&A bibliography style file (aa.bst)
\bibliography{Biedermann_S09} % your references in file: Yourfile.bib

\end{document}